\documentclass[twocolumn,aps,prb,floatfix,showpacs]{revtex4}
\usepackage{bm}
\usepackage{amssymb,amsmath}
\usepackage{dcolumn}
\usepackage{graphicx}
\usepackage{bbold}
\usepackage{booktabs}
\usepackage{color}
\usepackage[colorlinks,bookmarks=false,citecolor=blue,linkcolor=blue,urlcolor=blue]{hyperref}

\def\fuphat{\hat{f}_\uparrow}
\def\fdownhat{\hat{f}_\downarrow}
\def\nhat{\hat{n}}
\def\ahat{\hat{a}}
\def\bhat{\hat{b}}
\def\chat{\hat{c}}
\def\fhat{\hat{f}}

\def\alphahat{\hat{\alpha}}
\def\betahat{\hat{\beta}}
\def\phihat{\hat{\phi}}
\def\psihat{\hat{\psi}}
\def\r{{\bf r}}
\def\k{{\bf k}}
\def\R{{\bf R}}

\def\calU{{\cal U}}
\def\calE{{\cal E}}
\def\calK{{\cal K}}
\def\udel{\underline{\delta}}

\begin{document}
\title{Canonical representation for electrons and its application to the Hubbard model}
\author{Brijesh Kumar}
 \email{bkumar@mail.jnu.ac.in}
 \affiliation{School of Physical Sciences, Jawaharlal Nehru University, New Delhi 110067, India.}

\date{\today}%{December 19, 2007}

\begin{abstract}
A new representation for electrons is introduced, in which the electron operators are written in terms of a spinless fermion and the Pauli operators. This representation is canonical, invertible and constraint-free. Importantly, it simplifies the Hubbard interaction. On a bipartite lattice, the Hubbard model is reduced to a form in which the exchange interaction emerges simply by decoupling the Pauli subsystem from the spinless fermion bath. This exchange correctly reproduces the large $U$ superexchange. 
Also derived, for $U=\pm\infty$, is the Hamiltonian to study Nagaoka ferromagnetism.  In this representation, the infinite-$U$ Hubbard problem becomes elegant and easier to handle. Interestingly, the ferromagnetism in Hubbard model is found to be related to the gauge invariance of the spinless fermions. Generalization of this representation for the multicomponent fermions, a new representation for bosons, the notion of a `soft-core' fermion, and some interesting unitary transformations are introduced and discussed in the appendices.
\end{abstract}

\maketitle
%%%%%%%%%%%%
\section{Introduction}
Strongly correlated electrons represent a large class of physical systems exhibiting a variety of novel phenomena~\cite{Fazekas,TMO_Book}. The transition metal oxides, such as the high-T$_C$ superconducting cuprates~\cite{highTc,highTc_PWA,highTc_TVR,highTc_Lee}, or the colossal magneto-resistive manganites~\cite{manganites_review}, provide some of the most famous and actively investigated materials of this kind. The common feature in all such materials is the dominant presence of local repulsion between electrons, with manifestly non-trivial consequences, such as the antiferromagnetic-insulating behavior at half-filling (the Mott insulators), which can not be understood from the band-theory of solids~\cite{MITransition_Imada,PWA_super2}. Physics of these systems is intrinsically non-perturbative, and hence difficult. The Hubbard model~\cite{HubbardModel_1, HubbardModel_2, HubbardModel_3,Montorsi,Fazekas}, which describes a system of tight-binding electrons with on-site repulsion, allows the simplest meaningful description of the strongly correlated electrons. Since the local repulsion $U$ (also called the Hubbard interaction) and the hopping $t$ (assuming nearest-neighbor) have opposing effects on the electrons, the ground state of the Hubbard model can be metallic/insulating, magnetic/non-magnetic or something else (yet unexplored), depending upon the ratio $t/U$, the electron density $n_e$, and the geometry of the lattice. 

For sufficiently large $U$ at half-filling ($n_e=1$), the Hubbard model is known to systematically produce the antiferromagnetic exchange interaction and the insulating ground state~\cite{PWA_super2}.   As $t/U$ is increased, the insulating phase undergoes a quantum transition to a metallic phase. Though there have been many elaborate studies on Hubbard model~\cite{MITransition_Imada,DMFT_review}, there is no final verdict as yet on the nature of this metal-insulator transition (except in one dimension where the exact ground state is insulating for any non-zero $U$~\cite{Lieb_Wu}). Also, the search for a superconducting ground state in purely repulsive lattice models~\cite{OrderByProjection_BSS, OrderByProjection_KS, dwave_Hubbard_Jarrell} has become very important since the discovery of superconductivity in cuprates and other strongly correlated materials, but with no definite answer. Present understanding of the Hubbard model is far from being complete, and has been a subject of great research interest~\cite{Montorsi}. The main difficulty lies in the handling of the Hubbard interaction. This has led to the development of numerous approximate methods. The Gutzwiller projected variational wavefunctions~\cite{Gutzwiller,Brinkman_Rice},  the dynamical mean-field theory~\cite{DMFT_review},  the slave-boson (fermion) approach~\cite{Kotliar,Zou_PWA} or the cluster approximations~\cite{Cluster_review} are a few important ones to mention. Exact numerical diagonalization has also been used to study Hubbard model on small lattices~\cite{Dagotto}. The motivation to search for newer and more effective schemes of investigation, for a better understanding of the Hubbard model (and the strongly correlated systems in general), is therefore very strong.

In this paper, we present a new canonical representation for electrons, and demonstrate its usefulness for the Hubbard model. The basic idea is to identify the independent physical attributes of an electron, in terms of which the electron can be described as a composite. The two attributes of an electron are its {\em spin} and the {\em Fermi}-ness. The {\em spin} defines the two-level structure of the electron which can be described using Pauli operators. The {\em Fermi} attribute, which assigns the correct quantum statistics (anticommutation) to the electron, is described in terms of a spinless fermion. In our description, these two canonical degrees of freedom are combined to make the electron. An important feature of this representation is that it transforms the Hubbard interaction (the particle-hole symmetric form) into the chemical potential for the spinless fermions. The simplification thus achieved further encourages us to investigate the Hubbard model in this representation. We discuss two very important cases, that is the superexchange for finite $U$ and the Nagaoka ferromagnetism~\cite{Nagaoka} for infinite $U$, to highlight the physical usefulness of our representation in describing strongly correlated electrons. The paper is organized as follows. In the next section, we introduce the representation, discuss its canonical nature, and derive the inverse representation.  In section~\ref{sec:Hubbard_ne1}, we discuss the Hubbard model. On a bipartite lattice, we reduce the model to a nice form in which a part of the hopping explicitly generates the exchange interaction in a simple decoupling scheme. For large $U$ at half-filling, this exchange interaction correctly reproduces  Anderson's superexchange~\cite{PWA_super2}. In section~\ref{sec:Nagaoka}, we present the Hamiltonian formulation of the Nagaoka ferromagnetism. In the appendices, we discuss a few related but distinct topics. First, we generalize this representation for $N$-component fermions in appendix~\ref{sec:multicomponent}. Next, we write down a new representation for the bosons. In appendix~\ref{sec:Majorana_unitary}, we introduce a unitary transformation with Majorana gauge, and {\em derive} the Jordan-Wigner transformation~\cite{JW, Mattis}. Finally, we introduce the notion of a `soft-core fermion', and discuss its basic properties. 
%%%%%
\section{\label{sec:new_rep} New Representation for Electrons}
\begin{table}
\caption{\label{tab:states} Local Hilbert spaces and the operators of the electron, spinless fermion and a Pauli (two-level) system. Here, $|0\rangle$ and $|\o\rangle$ denote the vacua of the electron and the spinless fermion, respectively. Operators $\fhat^{ }_s$ and $\ahat$ satisfy the canonical Fermi algebra, while $\sigma^z$ and $\sigma^\pm$ are the Pauli operators.}
\begin{ruledtabular}
\begin{tabular}{l||c||c}
& Basis & Operators\\ \hline\hline && \\
Electron & $\{|0\rangle, |\uparrow\rangle, |\downarrow\rangle, |\uparrow\downarrow\rangle \}$&
$\fhat^{ }_s$, $\fhat^\dag_s$;~~$s=\uparrow, \downarrow$\\                                
&& \\
&$\begin{array}{rl} 
\fhat^{ }_s|0\rangle=& 0 \\
\fuphat^\dag |0\rangle=& |\uparrow\rangle \\
\fdownhat^\dag |0\rangle=& |\downarrow\rangle \\
\fuphat^\dag\fdownhat^\dag |0\rangle=& |\uparrow\downarrow\rangle 
\end{array}$ &
$\begin{array}{rl}  
\fuphat^\dag=&|\uparrow\rangle\langle0|+|\uparrow\downarrow\rangle\langle\downarrow|\\
\fdownhat^\dag=&|\downarrow\rangle\langle0|-|\uparrow\downarrow\rangle\langle\uparrow|
\end{array}$ \\
&& \\ \hline &&\\
Spinless &$\{ |\o\rangle, |1\rangle \}$&$\ahat$, $\ahat^\dag$\\
Fermion&&\\
&$\begin{array}{rl} \ahat|\o\rangle=& 0\\ \ahat^\dag|\o\rangle =& |1\rangle \end{array}$&
$\begin{array}{rl} \ahat^\dag = &|1\rangle\langle\o| \end{array}$\\
&& \\ \hline &&\\
Pauli &$\{|+\rangle,|-\rangle\}$&$\sigma^\pm$, $\sigma^z$ \\
System&&\\
&$\begin{array}{rl} \sigma^-|-\rangle=&0\\ \sigma^+|-\rangle=&|+\rangle \end{array}$&
$\begin{array}{rl} \sigma^+=&|+\rangle\langle-| \\  \sigma^z=&|+\rangle\langle+| - |-\rangle\langle-| \end{array} $
\end{tabular}
\end{ruledtabular}
\end{table}

Consider the Hilbert space of electron on a single site. The four basis states which span this local electronic Hilbert space are tabulated in Table~\ref{tab:states}, where $|0\rangle$ denotes the empty site, and $\fuphat^\dag$ and $\fdownhat^\dag$ are the creation operators of $\uparrow$ and $\downarrow$ electrons, respectively. Also consider a spinless fermion and a two-level (Pauli) system, each having a two-dimensional Hilbert space (see Table~\ref{tab:states}). Let $\ahat$ and $\ahat^\dag$ be the annihilation and creation operators of the spinless fermion, and $\sigma^z$ and $\sigma^\pm$ be the usual Pauli operators describing the two-level system. Since the composite Hilbert space of a spinless fermion and a Pauli system is four-dimensional, we can define a one-to-one mapping between the basis states of this composite Hilbert space and that of the electron~\cite{fnote_Feng,fnote_Mattis_Nam}. In Table~\ref{tab:map}, we define one such mapping, in which the one-electron states correspond to having one spinless fermion, and the {\em empty} or {\em double}-occupancy electronic states correspond to the vacuum of the spinless fermion. For the one-electron states, $|+\rangle$ corresponds to $\uparrow$ spin, and $|-\rangle$ to $\downarrow$ spin of the electron.  Otherwise, they correspond respectively to the double-occupancy or empty state. This mapping, which is motivated by our idea of the electron as a composite of the {\em Fermi} and {\em Pauli} degrees of freedom, results in the following new representation for the electron operators.
\begin{eqnarray}
 \fuphat^\dag & = & \phihat~\sigma^+ \label{eq:rep_up}\\
 \fdownhat^\dag & = & i\frac{\psihat}{2} - \frac{\phihat}{2}\sigma^z \label{eq:rep_down}
\end{eqnarray}
Here, $\phihat=\ahat^\dag + \ahat$, and $i\psihat= \ahat^\dag -\ahat$. The operators $\phihat$ and $\psihat$ (called the Majorana or real fermions) are Hermitian, and satisfy the relations: $\phihat^2=\psihat^2=1$ and $\{\phihat,\psihat\}=0$, where $\{,\}$ denotes the anti-commutator. As usual, the spinless fermion operators commute with the Pauli operators. We have derived Eqs.~(\ref{eq:rep_up}) and~(\ref{eq:rep_down}) using the definition of $\fuphat^\dag$ and $\fdownhat^\dag$ in terms of the basis states (as given in Table~\ref{tab:states}).

\begin{table}
\caption{\label{tab:map} A mapping between the states of the electron and a composite system of a spinless fermion and a Pauli system.}
\begin{ruledtabular}
\begin{tabular}{lcr}
    $|0\rangle$ & $\longleftrightarrow$ & $|\mbox{\o}\rangle~ |-\rangle$ \\
    $|\uparrow\rangle$ & $\longleftrightarrow$ & $|1\rangle~ |+\rangle$ \\
    $|\downarrow\rangle$ & $\longleftrightarrow$ & $|1\rangle~ |-\rangle$ \\
    $|\uparrow\downarrow\rangle$ & $\longleftrightarrow$ & $|\mbox{\o}\rangle~ |+\rangle$
\end{tabular}
\end{ruledtabular}
\end{table}

In this representation, the electron number operators are given as: \( \nhat_\uparrow = (1 + \sigma^z)/2 \) and \( \nhat_\downarrow = \frac{1}{2} + \left(\frac{1}{2} - \nhat\right)\sigma^z \), where $\nhat = \ahat^\dag \ahat$. The physical spin, ${\bf S}$, of the electron, which is defined as $S^z = (\nhat_\uparrow - \nhat_\downarrow)/2$ and $S^+ = \fuphat^\dag\fdownhat$, takes the form: ${\bf S}=\nhat\vec{\sigma}/2$. Similarly, the local pairing (the pseudo-spin operators), ${\bf P}$ can be written as: ${\bf P} = (\nhat-1)\vec{\sigma}/2$, where $P^+ = \fuphat^\dag\fdownhat^\dag$ and $P^z = (\nhat_\uparrow + \nhat_\downarrow - 1)/2$. Also, \( {\bf S} + {\bf P} = \vec{\sigma}/2 \) and \( S_i P_j  = 0\), where $i,j = x,y,z$. Here, $\sigma^x=\sigma^+ + \sigma^-$ and $\sigma^y=-i(\sigma^+ - \sigma^-)$.
Physically speaking, $\vec{\sigma}$ represents the spin of electron in the presence of the spinless fermion. Otherwise, it represents local pairing. These forms (especially that of ${\bf S}$) are consistent with  our intuitive picture in which an electron is envisaged as a ``composite" object made up of the {\em Fermi} and the {\em spin} attributes. Furthermore, it is important to note that \( \left(\nhat_\uparrow - \frac{1}{2}\right)\left(\nhat_\downarrow - 
\frac{1}{2}\right) = \frac{1}{2}\left(\frac{1}{2} - \nhat\right)\), that is the Hubbard interaction between electrons transforms into the number operator of the spinless fermion. 

The new representation of the electron is canonical, constraint-free and invertible. It is canonical because the electron operators satisfy the usual anti-commutation relations. It is obviously constraint-free because there are no extra conditions to be satisfied by $\ahat$ or $\vec{\sigma}$ operators. This is a consequence of the one-to-one mapping between the two Hilbert spaces. Since there are no excess or lost states, there is no constraint. Also, the spinless fermion and the Pauli operators can be represented uniquely in terms of the electron operators. The inverse of the new representation is given as:
\begin{eqnarray}
\ahat^\dag & = & (1-\nhat_\uparrow)\fdownhat^\dag - \nhat_\uparrow\fdownhat \label{eq:inv_rep_a}\\
\sigma^z & = & 2 \nhat_\uparrow - 1 \label{eq:inv_rep_sigmaz}\\
\sigma^+ & = & \fuphat^\dag\left(\fdownhat^\dag + \fdownhat\right) \label{eq:inv_rep_sigmaplus}
\end{eqnarray}
The inverse representation is also consistent, because the fermion operator, $\ahat$, satisfies the anti-commutation relation and the Pauli operators satisfy the angular momentum algebra. The Pauli operators also satisfy the anticommutation relation: $\{\sigma^i,\sigma^j\}=2\delta_{ij}$ for $i,j=x,y,z$. And of course, $\ahat$ and  $\ahat^\dag$ commute with $\vec{\sigma}$.

The inverse representation gives us more insight into the nature of the new objects. Equations~(\ref{eq:inv_rep_sigmaz}) and~(\ref{eq:inv_rep_sigmaplus}) are same as the Majorana fermion representation of the spin-1/2 operators~\cite{Shastry}. The spinless fermion [Eq.~(\ref{eq:inv_rep_a})] is effectively a Gutzwiller-projected electron whose creation is equivalent to the creation or annihilation of $\downarrow$ electron, subject to the absence or presence of $\uparrow$ electron respectively. It is also like the Bogoliubov quasiparticle in the BCS theory as it mixes a particle with a hole. Surprisingly, the inverse representation has   more familiar appearance, and hence, it is easier to appreciate. We can generate many equivalent forms of our representation by making suitable unitary rotations. This gives us the freedom of choosing a form that may serve us better for a given problem. For example, the unitary operator, ${\cal U}=1-2\nhat_\uparrow\nhat_\downarrow$, transforms Eqs.~(\ref{eq:inv_rep_a}-\ref{eq:inv_rep_sigmaplus}) as follows. 
\begin{eqnarray}
{\cal U}\ahat^\dag{\cal U}^\dag &=& \bhat^\dag=(1-\nhat_\uparrow)\fdownhat^\dag+\nhat_\uparrow\fdownhat \label{eq:rotate_a2b}\\
{\cal U}\sigma^+{\cal U}^\dag &=& \tau^+=-\fuphat^\dag(\fdownhat^\dag-\fdownhat)\label{eq:rotate_sigma2tau}
\end{eqnarray}
Or conversely, $\fuphat^\dag=i\psihat_b\tau^+$, and $\fdownhat^\dag=(\phihat_b - i\psihat_b~\tau^z)/2$, where $\phihat_b=\bhat^\dag+\bhat$ and $i\psihat_b=\bhat^\dag-\bhat$. We will use this and the original form of the representation to conveniently describe the Hubbard model on a bipartite lattice.
%%%%%%%
\section{\label{sec:Hubbard_ne1} Hubbard model at half-filling}
Consider the Hubbard model with nearest-neighbor hopping on a lattice. Its Hamiltonian in terms of the electron operators can be written as:
\begin{eqnarray}
H&=&-\frac{t}{2}\sum_{\r,\udel}\sum_{s=\uparrow,\downarrow} \left(\hat{f}^\dag_{\r,s} \hat{f}^{ }_{\r+\udel,s} + \hat{f}^\dag_{\r+\udel,s} \hat{f}^{ }_{\r,s} \right) \nonumber\\
 & &+U\sum_\r \left(\nhat_{\r\uparrow}-\frac{1}{2}\right)\left(\nhat_{\r\downarrow} - \frac{1}{2}\right) \label{eq:hubbard_model}
\end{eqnarray}
where $\r$ is summed over all the sites of a lattice, and $\udel$ is summed over the nearest-neighbors of a site. The electron hopping is denoted by $t$, and $U$ is the Hubbard interaction. In Eq.~(\ref{eq:hubbard_model}), the factor of $1/2$ in $t$ is there to undo the double-counting of the hopping due to the explicit use of its Hermitian conjugate. Therefore, the $t$ we use is same as the one commonly reported. The $U$-term in Eq.~(\ref{eq:hubbard_model}) is written in the particle-hole symmetric form~\cite{fnote_ph_symmetry}. The Hubbard model in the new representation [using Eqs.~(\ref{eq:rep_up}) and~(\ref{eq:rep_down})] takes the following form.
\begin{eqnarray}
H&=&-\frac{U}{2}\sum_\r \ahat^\dag_\r \ahat_\r -\frac{t}{2}\sum_{\r,\udel}\left(\sigma_\r^+~\sigma_{\r+\udel}^- - h.c.\right)~\phihat_\r~\phihat_{\r+\udel} \nonumber\\
 & &-\frac{i t}{4}\sum_{\r,\udel}\left(\sigma_\r^z~\phihat_\r~\psihat_{\r+\udel} + \sigma_{\r+\udel}^z~\phihat_{\r+\udel}~\psihat_\r\right) +\frac{U}{4}L \label{eq:hubbard_new}
\end{eqnarray}
Here, $L$ is the total number of lattice sites. In this form, the Hubbard interaction has taken the role of chemical potential for the spinless fermions, and the hopping has transformed into a complicated spin-fermion interaction. The Hubbard model has thus turned into a problem of spin-1/2 objects immersed in a grand-canonical bath of spinless fermions, which are constantly being created and annihilated around an average fermion density decided by $U$. The total $\sigma_z$ is however conserved, which is same the conservation of the total number of $\uparrow$-electrons. 

Since the local gauge of the electron governs the flow of charge-current in a system, it is important to consider how the new objects respond to such a local gauge transformation. This will give us an idea of the way charge is carried by the new objects. Let $\fhat^\dag_{\r,s} \rightarrow e^{i\theta_\r}\fhat^\dag_{\r,s}$ be the gauge transformation on electrons, under which the spinless fermions and the Pauli operators transform as: 
\begin{eqnarray}
\ahat^\dag_\r &\rightarrow&  (1-\nhat^{ }_{\r\uparrow} )\fhat^\dag_{\r\downarrow} e^{i\theta_\r} - \nhat^{ }_{\r\uparrow}\fhat^{ }_{\r\downarrow}e^{-i\theta_\r} \label{eq:gauge_fermi}\\
\sigma^+_\r &\rightarrow& \fhat^\dag_{\r\uparrow}\fhat^\dag_{\r\downarrow} e^{i2\theta_\r} + \fhat^\dag_{\r\uparrow}\fhat^{ }_{\r\downarrow} \label{eq:gauge_Pauli}
\end{eqnarray}
This implies that a spinless fermion carries one unit of electron or hole charge (of $\downarrow$ electron, to be more specific), and $\sigma^+_\r$ carries a charge of two electrons or no charge at all. It can be understood by explicitly considering the action of $\ahat^\dag$ and $\sigma^+$ on the electronic states. For completeness, we also write the current density operator for the Hubbard model in terms of the spinless fermions and the Pauli operators, which is $\vec{{\cal J}}_\r = (iet/2)\sum_{\udel} \udel (\fhat^\dag_{\r,s}\fhat^{ }_{\r+\udel,s} - \fhat^\dag_{\r+\udel,s}\fhat^{ }_{\r,s}) = (iet/4)\sum_{\udel}\udel [\psihat_\r\psihat_{\r+\udel} + \phihat_\r\phihat_{\r+\udel} (\vec{\sigma}_\r\cdot\vec{\sigma}_{\r+\udel})]$. Interestingly, the current operator has nicer form as compared to hopping in the new representation, and it also reveals the exchange interaction. At this point, it may be mentioned that Eq.~(\ref{eq:rotate_a2b}) is same as Eq.~(\ref{eq:gauge_fermi}) (with  some fine tuning) for $\theta_\r=\pi/2$. In fact, the ${\cal U}$ in Eq.~(\ref{eq:rotate_a2b}) was suggested by this observation.

On a bipartite lattice, Eq.~(\ref{eq:hubbard_new}) can be transformed to a more elegant form by making a suitable unitary rotation on the operators on one sublattice. Let $A$ and $B$ denote the two sublattices. We write the electron operators on $A$-sublattice, in terms of $\ahat$, $\ahat^\dag$ and $\vec{\sigma}$, as $\fuphat^\dag=\phihat_a\sigma^+$ and $\fdownhat^\dag=(i\psihat_a - \phihat_a\sigma^z)/2$. The electrons on $B$-sublattice are described in terms of $\bhat$, $\bhat^\dag$ and $\vec{\tau}$ such that $\fuphat^\dag = i\psihat_b\tau^+$ and $\fdownhat^\dag=(\phihat_b - i\psihat_b\tau^z)/2$.  The representation on $A$-sublattice is in the original form [as in Eqs.~(\ref{eq:rep_up}) and~(\ref{eq:rep_down})], while on $B$-sublattice, it is a unitary-transformed version of the same [see Eqs.~(\ref{eq:rotate_a2b}) and~(\ref{eq:rotate_sigma2tau})]. The Hubbard model on a bipartite lattice can therefore be written as:
\begin{eqnarray}
H &=&\frac{U}{4}L -\frac{U}{2}\left[\sum_{\R\in A}\ahat_\R^\dag\ahat^{ }_\R + \sum_{\R\in B}\bhat^\dag_\R\bhat^{ }_\R\right]  \label{eq:hubbard_bipartite}\\
& &-\frac{it}{2}\sum_{\R\in A}\sum_{\udel}\left[ \psihat^{ }_{a,\R}\phihat^{ }_{b,\R+\udel} + \psihat^{ }_{b,\R+\udel}\phihat^{ }_{a,\R}\left(\vec{\sigma}^
{ }_\R\cdot\vec{\tau}_{\R+\udel}\right)\right]\nonumber
\end{eqnarray}
This form of $H$ is neat and suggestive~\cite{fnote_Hubbard1}. It presents the {\em dual} role of hopping very clearly.  While a part of the hopping contributes purely to the dynamics of the spinless fermion bath, the remaining part generates an $SU(2)$ invariant exchange interaction~\cite{fnote_anisotropy}, whose strength is determined by the fluctuations in the bath. 

To gain further insight, we discuss the Hubbard model in a decoupled form: $H=H^{ }_{Fermi} + H^{ }_{Pauli}$, where
\begin{eqnarray}
H^{ }_{Fermi} &=& -\frac{U}{2}\sum_{\R\in A}\ahat_\R^\dag\ahat^{ }_\R + \frac{U}{2}\sum_{\R\in B}\bhat^{ }_\R\bhat^\dag_\R  \label{eq:H_fermi}\\
&& -t\sum_{\R\in A}\sum_{\udel} \left(\ahat^\dag_\R~\bhat^\dag_{\R+\udel} + \bhat^{ }_{\R+\udel}~\ahat^{ }_\R\right) \nonumber \\
&=&-\frac{U}{2}\sum_\k~\left(\ahat^\dag_\k\ahat^{ }_\k-\bhat^{ }_{\k}\bhat^\dag_{\k}\right)\label{eq:H_fermi_k}\\
& & -tz\sum_\k~\left(\gamma_\k~\ahat^\dag_\k\bhat^\dag_{-\k} + \gamma^*_\k~\bhat_{-\k}\ahat_\k\right)
\nonumber
\end{eqnarray}
and
\begin{equation}
H^{ }_{Pauli} = \frac{t\zeta}{2}\sum_{\R\in A}\sum_{\udel}~\left( \vec{\sigma}^{ }_\R\cdot\vec{\tau}^{ }_{\R+\udel} +1\right)\label{eq:H_Pauli}
\end{equation}
This decoupled from is derived in the following way. First, we rewrite $\vec{\sigma}^{ }_\R\cdot\vec{\tau}^{ }_{\R+\udel}$ in Eq.~(\ref{eq:hubbard_bipartite}) as $-1+(1+\vec{\sigma}^{ }_\R\cdot\vec{\tau}^{ }_{\R+\udel})$, and include $-\psihat^{ }_{b,\R+\udel}\phihat^{ }_{a,\R}$ resulting from this into the purely spinless fermion part of $H$, which is now called $H_{Fermi}$. Then, replace $\psihat^{ }_{b,\R+\udel}\phihat^{ }_{a,\R}$ [which is in multiplication with $(1+\vec{\sigma}^{ }_\R\cdot\vec{\tau}^{ }_{\R+\udel})$] by its expectation value in the ground state of $H_{Fermi}$. The resultant involving only Pauli operators is called $H_{Pauli}$. Thus, the bath of spinless fermions in the decoupled picture is described by $H_{Fermi}$, and $H_{Pauli}$ is the exchange Hamiltonian, where $(\vec{\sigma}_\R\cdot\vec{\tau}_{\R+\udel} +1)/2$ is the Dirac-Heisenberg exchange operator on a nearest-neighbor bond. Note that the constant term $UL/4$ has been absorbed in rewriting $\bhat^\dag_\R\bhat^{ }_\R$ as $\bhat^{ }_\R\bhat^\dag_\R$.

The validity of this decoupling can be tested against the well-known superexchange interaction, $4t^2/U$, in the limit of large $U$. The strength of the exchange interaction, $J$, is given as: $J=2t\zeta$, where $\zeta=-i\langle \psihat_{b,\R+\udel}\phihat_{a,\R}\rangle$ (uniform average on each nearest-neighbor bond) is determined purely by $H_{Fermi}$. Equation~(\ref{eq:H_fermi_k}) is the momentum space representation of $H_{Fermi}$, where $\gamma_\k=\frac{1}{z}\sum_{\udel} e^{i\k \cdot \udel}= |\gamma_\k| e^{i\lambda_\k}$, and $z$ is the nearest-neighbor coordination. Since $\gamma_\k^* = \gamma_{-\k}$, therefore, $|\gamma_{-\k}|=|\gamma_\k|$ and $\lambda_{-\k}=-\lambda_\k$. Here, the wavevector $\k$ lies in the sublattice Brillouin zone, and the Fourier transformation of the spinless fermions is defined as: \( \ahat^{ }_\R = \sqrt{\frac{2}{L}}\sum_\k e^{i\k\cdot\R}\ahat_\k \), and similarly for $\bhat_\k$. The $H_{Fermi}$ is diagonalized by applying two unitary rotations in following way.
\begin{eqnarray}
\tilde{H}_{Fermi} &=& \calU_\theta~\calU_\lambda~H_{fermi}~
\calU^\dag_\lambda~\calU^\dag_\theta \label{eq:H_fermi_diagonal}\\
&=& \sum_\k\calE_\k\left(1- \ahat^\dag_\k\ahat_\k - \bhat^\dag_\k\bhat_\k\right)\nonumber
\end{eqnarray}
The unitary operators in Eq.~(\ref{eq:H_fermi_diagonal}) are defined as: $\calU_\lambda=\prod_\k e^{i\lambda_\k\bhat^\dag_\k\bhat_\k}$, and $\calU_\theta=\prod_\k e^{\theta_\k(\ahat^\dag_\k\bhat^\dag_{-\k}-\bhat_{-\k}\ahat_\k)}$, where $\calU_\lambda$ absorbs the phase $\lambda_\k$ into $\bhat_\k$, and $\calU_\theta$ diagonalizes the $\ahat_\k$, $\bhat^\dag_{-\k}$ mixing. The quasi-particle dispersion, $\calE_\k=\sqrt{\frac{U^2}{4}+(zt|\gamma_\k|)^2}$, is positive, and $\tan{2\theta_\k} = zt|\gamma_\k|/(\frac{U}{2})$.
\begin{figure}
   \centering
   \includegraphics[width=8.5cm]{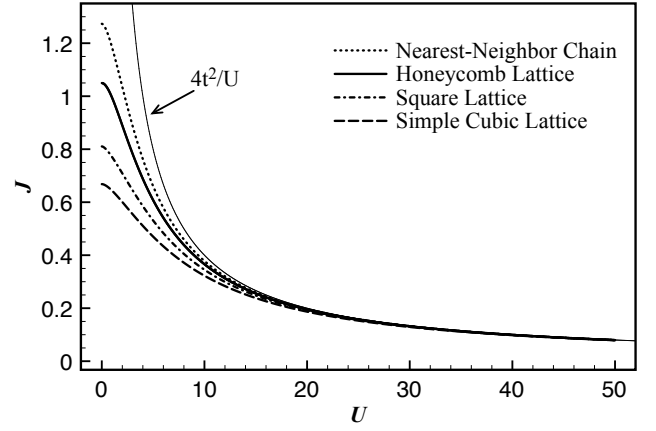} 
   \caption{Antiferromagnetic exchange $J$ as a function of $U$. The superexchange interaction, $4t^2/U$  ($t=1$ in our calculation), is also plotted for comparison. For large $U$, all $J\rightarrow 4t^2/U$.}
   \label{fig:JU}
\end{figure}

\begin{figure}
   \centering
   \includegraphics[width=8.5cm]{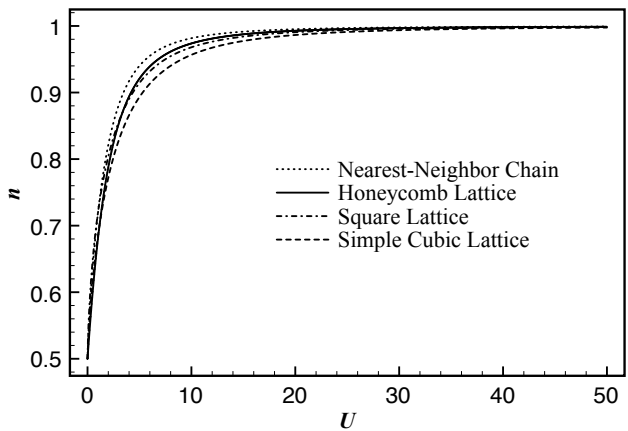} 
   \caption{Density of the spinless fermions, $n_a=n_b=n$ vs. $U$. For $U\gg 2zt$, $n\simeq 1$, and it is $0.5$ for $U=0$ (which is the correct value for the non-interacting case).}
   \label{fig:nU}
\end{figure}

The ground state of $\tilde{H}_{Fermi}$ is given by $\ahat^\dag_\k\ahat^{ }_\k=\bhat^\dag_\k\bhat^{ }_\k=1$, for all $\k$. We calculate $\zeta$ in the ground state of $H_{Fermi}$, and find that $J=\frac{4zt^2}{L}\sum_\k \frac{|\gamma_\k|^2}{\calE_\k}$, which is certainly positive. Hence, the exchange interaction in $H_{Pauli}$ is antiferromagnetic. Explicitly on a $D$-dimensional hypercubic lattice, where $z=2D$ and $\gamma_\k=\frac{1}{D}\sum_{i=1}^D\cos{k_i}$, we correctly retrieve Anderson's superexchange, $J\rightarrow 4t^2/|U|$, for $U\gg2zt$. In Fig.~\ref{fig:JU}, the calculated values of $J$ are plotted as a function of $U/t$, for a few different bipartite lattices. The densities of $a$ and $b$ type spinless fermions are calculated to be: $n_a=n_b=n=\frac{1}{2} + \frac{U}{2L}\sum_\k\frac{1}{\calE_\k} $.  For $U\gg 2zt$, $n \rightarrow 1$, and for $U\rightarrow 0$, $n \rightarrow \frac{1}{2}$ (see Fig.~\ref{fig:nU}). Both of these limiting cases give the expected values of the spinless fermion density. Since the local densities $n_{a,\R}$ and $n_{b,\R}$ also take the uniform value $n$, the electron density can be calculated as: $n_e=1+\frac{1-n}{L}\sum_\r\sigma^z_\r$, which is equal to $1$ when total $\sigma^z=0$ (that is, $\sum_A\sigma^z+\sum_B\tau^z=0$). Therefore, in the decoupled problem, the half-filling for electrons is achieved when the number of $\uparrow$ electrons is exactly $L/2$. Since the ground state of $H_{Pauli}$ corresponds to total $\sigma^z=0$, the half-filled condition for electrons is always satisfied in the ground state of the decoupled problem. Hence, the decoupling scheme discussed here is a consistent approximation, as it correctly yields the superexchange model of electron spins at half-filling for large $U$ (${\bf S}=\nhat\vec{\sigma}/2\simeq\vec{\sigma}/2$ for $\langle\nhat\rangle\simeq 1$). 

We close this section with the following observations. It is perhaps the first time that the superexchange interaction (which is truly a strong correlation effect) has been derived in a mean-field like approximation. This is to be contrasted with the usually known approaches which give rise to the superexchange only in the second order perturbation theory~\cite{PWA_super2,Zou_PWA}. Another point to be noted is that $J$ is an even function of $U$. Therefore, the electron pseudo-spins [${\bf P}=(1-\nhat)\vec{\sigma}/2\approx\vec{\sigma}/2$ for $\langle\nhat\rangle\rightarrow 0$] also interact via antiferromagnetic `superexchange' for large negative $U$. This can be deduced, without explicit calculation, from the particle-hole ($ph$) transformation on $H_{Fermi}$, which implies that $(n; U,t) \stackrel{ph}{\longleftrightarrow} (1-n;-U,t)$. The present approach also gives a reasonable picture for $U=0$. In this case, the ground state energy of the bath is exactly half of the energy of the half-filled Fermi-sea,  because the spinless fermions do not account for the electron spin. In principle, the remaining half could be retrieved from $H_{Pauli}$, if each nearest-neighbor bond were in the singlet state. It clearly points at the singlet character of the free-electron ground state. Also, the $J$ is linear in $t$ (unlike the superexchange $J$), and gives a measure of the exchange energy of the Fermi-sea. It appears to us that this representation may be able to describe the large and small $U$ limits in a unified way. We will present more studies on Hubbard model in a future work. 
 
%%%%%%%%%%%%%%%%%%
\section{\label{sec:Nagaoka} $|U|=\infty$: Nagaoka ferromagnetism}
We now address the case of infinite $U$ Hubbard model. It turns out that taking $U\rightarrow\pm\infty$ limit is straightforward in the new representation. Since $U$ is the chemical potential for spinless fermions, the infinite $|U|$ completely suppresses the spontaneous fluctuations in the number of spinless fermions. That is, in the limit $|U|\rightarrow\infty$, the grand canonical bath of spinless fermions becomes canonical. For positive-infinite $U$, the Hubbard model on square lattice is known to have  ferromagnetic ground state close to half-filling (Nagaoka ferromagnetism~\cite{Nagaoka}), which eventually becomes unstable around $n_e\sim1\pm \delta_c$ (early estimate of the critical (hole) concentration is $\delta_c\sim 0.49$~\cite{Nagaoka_SKA}; later works suggest $\delta_c \sim 0.25$~\cite{Linden_Edwards}). There have been many studies on the existence and stability of the Nagaoka ferromagnetism on different types of lattices. Presently, we discuss this phenomenon in our representation, for positive as well as negative infinite-$U$ Hubbard model on bipartite lattices. The $|U|=\infty$ Hamiltonian in this case can be written as:
\begin{eqnarray}
H_\infty = -t\sum_{\R,\udel}\frac{\left(1+\vec{\sigma}_\R\cdot\vec{\tau}_{\R+\udel} \right)}{2}\left[\ahat^\dag_\R\bhat^{ }_{\R+\udel} + \bhat^\dag_{\R+\udel}\ahat^{ }_\R\right] 
\label{eq:H_infty}
\end{eqnarray}
where the hopping of the spinless fermions is exclusively determined by the exchange operator on each nearest-neighbor bond, in a way, similar to (but not same as) the Anderson-Hasegawa double-exchange~\cite{Hasegawa} in a ferromagnetic metal. 
Although $H_\infty$ is the same for both signs of infinite-$U$, the two limits correspond to different states of the spinless fermion bath. For $U=+\infty$, the bath is fully-filled ($n=1$), while it is completely empty ($n=0$) for $U=-\infty$. Both of these cases are dead, as there are no fermions which can gain energy by hopping. Also, there is $2^L$-fold degeneracy due to $L$ independent Pauli objects. In terms of electrons, the former ($U=+\infty$) corresponds to the localized electrons at half-filling, while in the latter case, an arbitrary site in a configuration is either empty or doubly occupied. In order to discuss interesting physical possibilities, we `externally' populate (or depopulate) the fermion bath, by employing the chemical potential, $\mu$, for electrons. That is, we consider $\calK_\infty = H_\infty-\mu (N_\uparrow + N_\downarrow)$, where $N_\uparrow + N_\downarrow=\sum_{\R\in A}[1+(1-\ahat^\dag_\R\ahat^{ }_\R)\sigma^z_\R] +\sum_{\R\in B}[1+(1-\bhat^\dag_\R\bhat^{ }_\R)\tau^z_\R]$. Below, we shall discuss the existence and stability of the ferromagnetic ground state in $\calK_\infty$. We will see that the present approach is considerably simpler as compared to the usual Gutzwiller projection based methods. 
 
First we discuss $H_\infty$ on just two sites. The exchange operator on a bond has eigenvalue $+1$ for the triplet and $-1$ for the singlet state of the two-level objects. Although the complete eigenstates of $H_\infty$ in the two cases are degenerate, they physically presents two distinct possibilities in the ground state. For a fully polarized triplet $|-,-\rangle$, the ground state of the two-site problem is: $( |1,\o\rangle+|\o,1\rangle)\otimes |-,-\rangle\equiv |\downarrow,0\rangle+|0,\downarrow\rangle$ (ignoring the normalization factors). This is a trivial `ferromagnetic-metallic' state with $\downarrow$ magnetization.  Similarly for $|+,+\rangle$, it is $|\uparrow,\uparrow\downarrow\rangle + |\uparrow\downarrow,\uparrow\rangle$, which is also ferromagnetic (but with respect to the fully-filled electron configuration). For the singlet state, $(|+,-\rangle-|-,+\rangle)$, the ground state is: $(|1,\o\rangle - |\o,1\rangle)\otimes(|+,-\rangle-|-,+\rangle) \equiv |\uparrow,0\rangle + |0,\uparrow\rangle + |\uparrow\downarrow,\downarrow\rangle + |\downarrow,\uparrow\downarrow\rangle$, which is like a `non-magnetic' metallic state. Also, in the singlet case, $n_\uparrow=n_\downarrow=1/2$ and ${\bf S}=0$. This simple analysis suggests two qualitatively different possibilities in the ground state of $H_\infty$.

Now we study $\calK_\infty$ on the full lattice. We explicitly consider the case of $U=-\infty$ (that is,  gradually fill the empty bath). The results for $U=+\infty$ can be exactly inferred from the particle-hole transformation on the spinless fermions, which states that the ground state for $+ve$ $U$ for a given density $n$, is same as that of $-ve$ $U$ for  $1-n$, for a fixed $t$ and $m$  [where $m=(\sum_A\sigma^z+\sum_B\tau^z)/L$]. To establish the existence of ferromagnetic-metallic ground state for $U=-\infty$, consider one spinless fermion in the empty bath. The (only) energy, that it gains by hopping, can be maximized by having maximum and uniform hopping through each bond. While the fully-polarized `ferromagnetic' state of the Pauli subsystem fulfills this requirement, the same can not be achieved in any other state, for instance, a configuration of bond-singlets or any deviation with respect to the fully polarized state. The fully polarized Pauli subsystem therefore emerges as the only choice which helps the fermion in gaining maximum kinetic energy.  The same will be true for a finite density of the spinless fermions, until the `ferromagnetic' state of the Pauli subsystem becomes unstable. In terms of electrons, the ground state of a finite $n$ fermion bath together with `$-$' polarized Pauli subsystem, corresponds to a $\downarrow$ polarized ferromagnetic metal with the electron density, $n_e=n$. Also, for the `$+$' polarized Pauli subsystem, the electronic ground state is ferromagnetic, but from the fully-filled side of the electron density ($n_e=2-n$). This analysis further implies that the ground state of $U=+\infty$ Hubbard model is also ferromagnetic-metallic, as discovered by Nagaoka, around half-filling ($n_e=1\pm n$, for `$+$' or `$-$' polarization respectively). Below we discuss the instability of the ferromagnetic state. It is generally a hard problem. But in our framework, it actually becomes a much simpler problem.

\begin{figure}
   \centering
   \includegraphics[width=8.5cm]{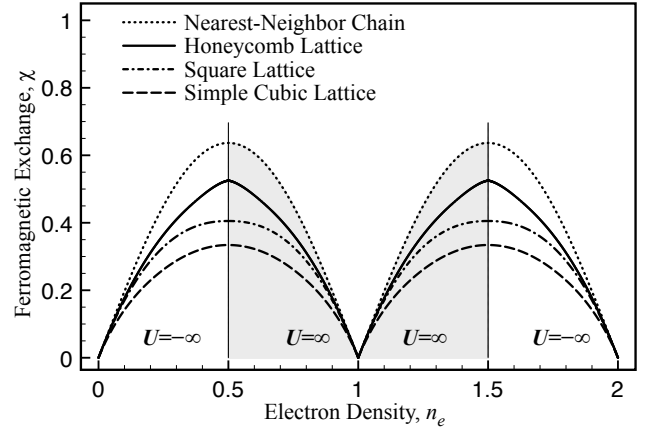} 
   \caption{Metallic ferromagnetism in the infinite-$U$ Hubbard model. The shaded region between $n_e=0.5$ and $1.5$ is the Nagaoka ferromagnetic state for $U=+\infty$. Note the complementary existence (in the sense of electron filling) of the Nagaoka ferromagnetism for $U=+\infty$ and $-\infty$.}
   \label{fig:Nagaoka}
\end{figure}

As it appears from the structure of $H_\infty$, the spinless fermions are happy to live with ferromagnetic Pauli subsystem for any density of electrons. But the Pauli subsystem itself is not comfortable with the change in $n_e$, because $\mu$ acts as  an `external' field for the Pauli operators. This induces `flips' in the fully-polarized Pauli subsystem for sufficiently populated fermion bath, thereby causing the instability of the Nagaoka state. To work this out, we write the Pauli operators in terms of $\alphahat$ and $\betahat$, the two hard-core bosons, such that: \( \sigma^z=-1+2\alphahat^\dag\alphahat \), $\sigma^+=\alphahat^\dag$ and \(\tau^z=-1+2\betahat^\dag\betahat \), $\tau^+=\betahat^\dag$. The creation of a hard-core boson amounts to flipping $|-\rangle$ on the respective site to $|+\rangle$, like a `magnon'. Rewriting $\calK_\infty$ in terms of $\alphahat$ and $\betahat$ separates the ferromagnetic part from the terms which cause flips, that is: $\calK_\infty=H^{ }_{FM} + H^{ }_{flips}$, where
\begin{eqnarray}
H^{ }_{FM} &=& -t\sum_{\R\in A}\sum_{\udel} (\ahat^\dag_\R\bhat^{ }_{\R+\udel} + \bhat^\dag_{\R+\udel}\ahat^{ }_\R) \label{eq:H_FM}\\
&& -\mu\sum_{\R\in A}\ahat^\dag_\R\ahat^{ }_\R -\mu\sum_{\R\in B}\bhat^\dag_\R\bhat^{ }_\R \nonumber
\end{eqnarray}
and
\begin{eqnarray}
H^{ }_{flips} &=& t\sum_{\R\in A}\sum_{\udel} \left(\ahat^\dag_\R\bhat^{ }_{\R+\udel} + \bhat^\dag_{\R+\udel}\ahat_\R\right)\times \label{eq:H_flips}\\ 
& &\left[ (\alphahat^\dag_\R\alphahat^{ }_\R + \betahat^\dag_{\R+\udel}\betahat^{ }_{\R+\udel}) -  (\alphahat^\dag_\R\betahat^{ }_{\R+\udel} + \betahat^\dag_{\R+\udel}\alphahat^{ }_\R) \right. 
\nonumber \\
&& \left. - 2(\alphahat^\dag_\R\alphahat^{ }_\R)(\betahat^\dag_{\R+\udel}\betahat^{ }_{\R+\udel}) \right] \nonumber\\
&& -2\mu\sum_{\R\in A}(1-\ahat^\dag_\R\ahat^{ }_\R)\alphahat^\dag_\R\alphahat^{ }_\R \nonumber\\
&& -2\mu\sum_{\R\in B}(1-\bhat^\dag_\R\bhat^{ }_\R)\betahat^\dag_\R\betahat^{ }_\R \nonumber
\end{eqnarray}
In the ferromagnetic state, $H^{ }_{flips}$ is zero as all the two-level objects are in the same state, and only $H^{ }_{FM}$ accounts for the ground state for different $n_e$, where $n_e=n$. 

Instability of the ferromagnetic state can be determined by calculating the dispersion of exactly one `flip', in the ground state of $H^{ }_{FM}$ for different values of $\mu$. That is, we replace the fermion operator terms in $H^{ } _{flips}$ by their expectation values in the ground state of $H^{ }_{FM}$, and calculate the one-flip dispersion. That value of $n_e$, for which the minimum of this dispersion becomes negative, marks the instability for the Nagaoka state. By following the above prescription, we get the following equations for the spinless femion density, $n$, and the one-flip dispersion, $\epsilon_\k^\pm$.
\begin{eqnarray}
n &=& \frac{1}{L}\sum_\k \left[\Theta(-zt|\gamma_\k|-\mu) + \Theta(zt|\gamma_\k|-\mu)\right] \\
&& \nonumber \\
\chi &=& -\frac{2}{zt}\frac{\langle H_\infty\rangle}{L} \\
&=& \frac{2}{L}\sum_\k |\gamma_\k|\left[\Theta(-zt|\gamma_\k|-\mu) - \Theta(zt|\gamma_\k|-\mu) \right] \nonumber\\
&& \nonumber \\
\epsilon^{\pm}_\k &=& zt\chi(1\pm |\gamma_\k|) - 2\mu (1-n)
\end{eqnarray}
Here, $\chi$ is the ferromagnetic ground state energy per nearest-neighbor bond (in units of $-t$), and $t\chi$ is always positive (for any sign of $t$). Since $zt\chi(1\pm|\gamma_\k|)\ge 0$ and $(1-n)$ is also positive, therefore, $\epsilon_\k^\pm$ is certainly non-negative for $\mu < 0$, which implies a stable ferromagnetic ground state. For $\mu=0^+$ however, $\epsilon^-_\k$ becomes negative at $\k=0$. Thus, the ferromagnetic ground state becomes unstable to the flips in the Pauli subsystem, and there occurs a first order transition in the ground state. The stable ferromagnetic-metallic ground state ($\mu<0$) corresponds to $n_e=n<1/2$, and the instability exactly occurs at $n_e=1/2$ ($\mu=0$). The same analysis for `$+$' polarized case gives the stable Nagaoka state for $n_e=2-n$, where $0< n < 1/2$. Using the particle-hole transformation for spinless fermions, we further deduce that the Nagaoka state for $U=+\infty$ exists for $1/2 < n_e < 1$ (hole doping) and for $1 <n_e < 3/2$ (electron doping). These results are presented in Fig.~\ref{fig:Nagaoka}. 

This simple analysis of a difficult problem is very encouraging. It gives the essential physics, that is the ferromagnetic ground state and its instability, in a very straightforward manner. We could show the existence of Nagaoka state for  both $U=+\infty$ as well as $-\infty$ Hubbard model. Of course, we would like to investigate (more accurately) whether the critical $n$ is exactly 0.5 or slightly different. But $n=0.5$ appears to be a special fermion density for bipartite lattices (maximum ferromagnetic exchange; also, corresponds to a non-magnetic metallic ground state). Further investigations of the Nagaoka ferromagnetism for {\em finite} $U$ case, and on non-bipartite lattices will be discussed elsewhere. Finally, an important point to note in the present discussion is that the ferromagnetism in Hubbard model is strictly related to the invariance of the spinless fermions under a global gauge transformation, while the antiferromagnetic/non-magnetic states break this gauge invariance. This is a novel point of view for the strongly correlated electrons.
%%%%
\section{Summary}
\begin{itemize}
\item A new canonical and invertible representation for the electrons, in terms of a spinless fermion and the Pauli operators, is presented. It is further generalized for the $N$-component fermions in appendix~\ref{sec:multicomponent}. 
\item The Hubbard interaction in this representation acts as the chemical potential for spinless fermions. On a bipartite lattice, the Anderson's superexchange for large and positive (as well as negative) $U$ is shown to correctly emerges from the spinless fermion bath, within a simple decoupling scheme.
\item A Hamiltonian is derived for studying Nagaoka ferromagnetism in $|U|=\infty$ Hubbard model. The existence of the ferromagnetic-metallic state and its stability is discussed. It is pointed out that the ferromagnetism in Hubbard model  respects the global gauge invariance for the spinless fermions, while the antiferromagnetism necessarily violates it.
\item The meaningful analysis of the two important cases of the Hubbard model is very encouraging for the new representation, and demonstrates its usefulness for studying strongly correlated electrons.  
\item A representation for bosons is discussed in appendix~\ref{sec:bose_rep}. The Jordan-Wigner transformation is derived in appendix~\ref{sec:Majorana_unitary} by introducing a unitary transformation with a Majorana fermion as gauge. Finally, the notion of a soft-core fermion is introduced in appendix~\ref{sec:soft-core}.
\end{itemize}
\begin{acknowledgments}
I am deeply indebted to Professor Sriram Shastry for critical discussions and constant encouragement.
\end{acknowledgments}
%%%%%%
\appendix
\section{\label{sec:multicomponent} Generalization to multicomponent fermions} 
Let $\fhat^\dag_i$ be $i^{th}$ component of an $N$-component fermion operator. Its Hilbert space is $2^N$ dimensional. In terms of a spinless fermion and $N-1$ two-level objects, both the equality of the Hilbert space dimensions and the anti-commutativity can be attained consistently for the $N$-component fermion, by defining a representation in the following way.
\begin{eqnarray}
\fhat^\dag_1 &=& \phihat~\sigma_1^+ \label{eq:new_repN_1}\\
\fhat^\dag_i &=& \phihat~\prod_{j=1}^{i-1}\sigma_j^z~\sigma_i^+~~~~\mbox{for $i=2,\cdots, N-1$} \label{eq:new_repN_i}\\
\fhat^\dag_N &=& \frac{i\psihat}{2} -\frac{\phihat}{2}\prod_{i=1}^{N-1}\sigma_i^z \label{eq:new_repN_N}
\end{eqnarray}
The operators $\fhat^\dag_i$ (for $i=1,N-1$) anticommute with each other because $\sigma^+_i$ anticommutes with $\sigma^z_i$, and $\fhat_N^\dag$ anticommutes with the rest of $\fhat^\dag_i$ operators, because $\psihat$ anticommutes with $\phihat$ (in addition to the anticommutation of $\sigma^+_i$ and $\sigma^z_i$). The inverse of this representation [Eqs.~(\ref{eq:new_repN_1}-\ref{eq:new_repN_N})] is given as:
\begin{eqnarray}
\mbox{For} && i=1,N-2 \nonumber\\
\sigma_i^+ &=&\fhat_i^\dag~\prod_{j=i+1}^{N-1}(2\nhat_j-1)~\left(\fhat_N^\dag+\fhat_N\right) \label{eq:inv_repN_1}\\ 
\sigma_{N-1}^+ &=&\fhat^\dag_{N-1}~\left(\fhat^\dag_N+\fhat_N\right) \label{eq:inv_repN_2}\\
\ahat^\dag &=&\fhat^\dag_N\left[\frac{1-\prod_{i=1}^{N-1}(2\nhat_i-1)}{2}\right]-\nonumber\\ && \fhat_N\left[\frac{1+\prod_{i=1}^{N-1}(2\nhat_i-1)}{2}\right] \label{eq:inv_repN_N}
\end{eqnarray}
where $\nhat_i=\fhat^\dag_i\fhat_i$.
The similarity between Eqs.~(\ref{eq:inv_repN_1}-\ref{eq:inv_repN_2}) and the Jordan-Wigner transformation is notable, with the exception that the Jordan-Wigner does not have an extra $\left(\fhat^\dag_N+\fhat_N\right)$ operator [and no equivalent of Eq.~(\ref{eq:inv_repN_N})]. Moreover,  Eqs.~(\ref{eq:new_repN_1}) and~(\ref{eq:new_repN_i}) are similar to the inverse Jordan-Wigner transformation [without a $\phihat$ and Eq.~(\ref{eq:new_repN_N})]. One can generate many equivalent representations for the $N$-component fermion by making suitable unitary rotations~\cite{fnote_Nfermi}.

%%%%%%

\section{\label{sec:bose_rep} Representation for bosons}
Let $\ahat^\dag$ and $\ahat$ be the creation and annihilation operators of a canonical boson, acting on the Fock states, $\{| n \rangle,~n=0,1\dots,\infty\}$. The operator $\ahat^\dag$ can be written as: 
\begin{eqnarray}
\ahat^\dag &=& \sum_{n=0}^\infty \sqrt{n+1}~|n+1\rangle\langle n| \label{eq:bose1}\\
&=&\sum_{m=0}^\infty \sqrt{2m+1}~|2m+1\rangle\langle 2m|  \label{eq:bose2}\\
&& + \sum_{m=0}^\infty \sqrt{2(m+1)}~|2(m+1)\rangle\langle 2m+1|  \nonumber
\end{eqnarray}
where $n=2m$ in the first summation in Eq.~(\ref{eq:bose2}), and $n=2m+1$ for the second. This even-odd structure of the Fock states allows us to write a composite representation of the boson operators by defining the following mapping.
\begin{eqnarray}
|2m\rangle &:=& |m\rangle\otimes |e\rangle \label{eq:even} \\
|2m+1\rangle&:=& |m\rangle\otimes |o\rangle \label{eq:odd}
\end{eqnarray} 
The kets $|e\rangle$ and $|o\rangle$ stand respectively for the even and odd values of the original Fock state quantum number $n$, and the states $\{|m\rangle\}$ define a `new' Fock space. 
Substituting this mapping in Eq.~(\ref{eq:bose2}) results in the following representation of $\ahat^\dag$.
\begin{eqnarray}
\ahat^\dag &=&\sqrt{2\bhat^\dag\bhat+1}~\sigma^+ + \sqrt{2}~\bhat^\dag \sigma^- \label{eq:bose_rep}
\end{eqnarray}
Here, $\sigma^\pm$ are the Pauli operators defined in the Hilbert space of $|e\rangle$ and $|o\rangle$, such that $\sigma^+ = |o\rangle\langle e|$. Operators $\bhat^\dag$ and $\bhat$ are the new canonical bosons, acting in the Fock space spanned by $\{|m\rangle\}$, which commute with the Pauli operators. It is easy to check that Eq.~(\ref{eq:bose_rep}) satisfies the bosonic commutation relations. The harmonic oscillator in terms of the original bosons takes the following form in the new representation. 
\begin{equation}
\ahat^\dag\ahat+\frac{1}{2}= 2\left[\bhat^\dag\bhat +\frac{1}{2}\right] + \frac{1}{2}\sigma^z \label{eq:atom_field}
\end{equation}
The new form looks like a two-level atom in a quantized single-mode radiation field (without atom-field coupling), which conversely suggests that an atom-field problem can be exactly mapped to an equivalent field-only problem (a kind of `unification' of matter and radiation-field into a new radiation-field). Most direct and simple case of this unification, present in Eq.~(\ref{eq:atom_field}) itself, is when the frequency of radiation is twice the energy-difference between the two atomic-levels [see right-hand side of Eq.~(\ref{eq:atom_field})]. The presence of an atom-field interaction will further introduce non-trivial terms in the unified radiation field. The application of this representation to an atom-field interaction problem will be discussed elsewhere~\cite{atom_field_paper}. 

Let us write down the inverse of this representation. The Pauli operators, in terms of $\ahat$ and $\ahat^\dag$, are given as:
\begin{eqnarray}
\sigma^z &=& -\cos{\pi\ahat^\dag\ahat} :=-\hat{\chi} \label{eq:sigmaz_bose}\\
\sigma^+ &=& \frac{1-\hat{\chi}}{2}~\frac{1}{\sqrt{\ahat^\dag\ahat}}~\ahat^\dag~\frac{1+\hat{\chi}}{2} \label{eq:sigmap_bose}
\end{eqnarray}
Note that $(1\pm \hat{\chi})/2$ are the projection operators of the odd and even states respectively, in the original Fock space. Therefore, the meaning of Eq.~(\ref{eq:sigmap_bose}) is clear that it connects an even $|n\rangle$ to an odd $|n\rangle$ and annihilates the odd states, consistent with the definition of $\sigma^+$. Also, the Pauli operators in the above representation satisfy the necessary algebra. Interestingly, Eqs.~(\ref{eq:sigmaz_bose}) and~(\ref{eq:sigmap_bose}) also give us an unconstrained {\em new} bosonic representation for spin-1/2 operators. If we restrict the Fock space to $|0\rangle$ and $|1\rangle$ (hard-core constraint), then we retrieve the usual hard-core boson representation. Now to the boson $\bhat^\dag$ which, in terms of $\ahat$ and $\ahat^\dag$, is represented as:
\begin{eqnarray}
\bhat^\dag &=& \frac{1}{\sqrt{2}}\left[\frac{1-\hat{\chi}}{2}~\frac{1}{\sqrt{\ahat^\dag\ahat}}~\ahat^\dag\ahat^\dag~\frac{1-\hat{\chi}}{2}\right] +  \nonumber\\ &&
\frac{1}{\sqrt{2}}\left[\frac{1+\hat{\chi}}{2}~\ahat^\dag\ahat^\dag~\frac{1}{\sqrt{\ahat^\dag\ahat+1}}~\frac{1+\hat{\chi}}{2}\right] 
\label{eq:b_bose}
\end{eqnarray}
It is consistent with the fact that the number of $b$-bosons changes by 1 when the number of $a$-bosons changes by 2. Besides, it satisfies the bosonic commutations, and commutes with $\sigma^z$ and $\sigma^\pm$, as in Eqs.~(\ref{eq:sigmaz_bose}) and~(\ref{eq:sigmap_bose}). Hence, a canonical and invertible representation for bosons.
%%%%%%

\section{\label{sec:Majorana_unitary} Unitary transformations with `Majorana' gauge}
We now introduce a new kind of gauge transformation, and show how it generates the Jordan-Wigner representation (JW) for spin-1/2 operators. Usually, we know the JW as an intuitive mapping that was not derived. Here, we precisely do that: {\em derive it}. Consider $\sigma^+$ in terms of the electron operators [as in Eq.~(\ref{eq:inv_rep_sigmaplus})]. It is given as: $\sigma^+=\fhat^\dag_\uparrow\phihat^{ }_\downarrow$, where $\phihat^{ }_\downarrow=\fhat^\dag_\downarrow+\fhat^{ }_\downarrow$ is a Majorana fermion corresponding to $\downarrow$ electron. Since $\phihat^2_\downarrow=1$, therefore, \(e^{\pm i\frac{\pi}{2}\phihat^{ }_\downarrow} = \pm e^{i\frac{\pi}{2}}\phihat^{ }_\downarrow\), and $\sigma^+$ can be rewritten as:
\begin{equation}
\sigma^+ = \fhat^\dag_\uparrow e^{-i\frac{\pi}{2}(1-\phihat^{ }_\downarrow)} = e^{-i\frac{\pi}{2}(1+\phihat^{ }_\downarrow)}\fhat^\dag_\uparrow
\label{eq:sigmaplus_gauge}
\end{equation}
where the $\downarrow$ electron Majorana fermion appears in the gauge. This observation motivates us to define a suitable `Majorana' gauge-transformation such that the phase factor in Eq.~(\ref{eq:sigmaplus_gauge}) is completely absorbed into $\fhat^\dag_\uparrow$, thereby transforming $\sigma^+$ into $\fhat^\dag_\uparrow$. We easily define such a transformation as: 
\begin{eqnarray}
\calU(\phihat^{ }_\downarrow) &=& \exp{\left[-i\frac{\pi}{2}(1+\phihat^{ }_\downarrow) \nhat^{ }_\uparrow\right]} \label{eq:U_phi}\\
&=&(1-\nhat^{ }_\uparrow) -\phihat^{ }_\downarrow\nhat^{ }_\uparrow \nonumber
\end{eqnarray}
where $\nhat^{ }_\uparrow=\fhat^\dag_\uparrow\fhat^{ }_\uparrow$, and the expression in the second line has been derived from the first by using the properties: $\left(\frac{1+\phihat_\downarrow}{2}\right)^ 2= \frac{1+\phihat^{ }_\downarrow}{2}$ and $\nhat^2_\uparrow=\nhat^{ }_\uparrow$. Clearly, $\calU(\phihat^{ }_\downarrow)$ is both unitary as well as Hermitian. It is a simple extension of the usual gauge transformation, achieved by allowing a Majorana fermion to be the phase. One can show now that $\calU^\dag(\phihat^{ }_\downarrow)\sigma^+\calU(\phihat^{ }_\downarrow) = \fhat^\dag_\uparrow$, while $\sigma^z$ remains the same, that is $2\nhat_\uparrow-1$. This is nice but unusual, because under a `normal' unitary transformation, a canonical fermion transforms into another canonical fermion. But here, under the Majorana gauge transformation, a canonical fermion becomes a Pauli operator (hard-core boson) and vice-versa. This change in the canonical nature happens due the fact that the gauge itself is a real fermion, not just a number. Next, we discuss the consequence of applying the Majorana gauge transformation [as in Eq.~(\ref{eq:U_phi})] on a collection of spin-1/2 objects. 

Let there be $L$ number of spin-1/2 objects described in terms of the Pauli operators, $\sigma_l^+$, where the integer label $l$ goes from $1$ to $L$. In order to transform every $\sigma^+_l$ into a corresponding $\fhat^\dag_{l\uparrow}$, we must apply $\calU(\phihat^{ }_{l\downarrow})$ for each $l$. Since $\calU(\phihat^{ }_{l\downarrow})\calU(\phihat^{ }_{m\downarrow})\neq\calU(\phihat^{ }_{m\downarrow})\calU(\phihat^{ }_{l\downarrow})$, it is important to keep track of the order in which to make the local gauge transformations. We can do this by defining an ordered unitary operator, $\calU$, as follows.
\begin{equation}
\calU = \calU(\phihat^{ }_{L\downarrow})~\cdots~\calU(\phihat^{ }_{2\downarrow})~\calU(\phihat^{ }_{1\downarrow}) \label{eq:ordered_U}
\end{equation}
In principle, we could define $\calU$ in any order, but we just choose to work with the above arrangement. Now, we apply $\calU$ one by one on each $\sigma^+_l$, and get the following answer, which is the JW transformation.
\begin{eqnarray}
\calU^\dag\sigma_1^+\calU &=& \fhat^\dag_{1\uparrow} \\
\calU^\dag\sigma_l^+\calU &=& \left[\prod_{m=1}^{l-1}(1-2\nhat^{ }_{m\uparrow})\right] \fhat^\dag_{l\uparrow} \\ && \nonumber \\
 &&\forall~~l=2\dots L \nonumber
\end{eqnarray}
We have thus derived the JW representation for spin-1/2 operators by applying a unitary transformation (with Majorana gauge) on the Majorana representation of the same. The origin of the string of $(1-2\nhat^{ }_{l\uparrow})$ in the JW representation also becomes clear from the nature of $\calU$. For $\sigma^+_l$, the operators $\calU(\phihat^{ }_{L\downarrow})$ to $\calU(\phihat^{ }_{l+1\downarrow})$ do nothing, while $\calU(\phihat^{ }_{\downarrow})$ gives $\fhat^\dag_{l\uparrow}$. The rest which follow afterwards, in the given order, contribute a factor of $(1-2\nhat^{ }_\uparrow)$ each, because every $\calU(\phihat^{ }_{m\downarrow})$ ($\forall$ $m<l$) that commutes past $\fhat^\dag_{l\uparrow}$ becomes $\calU(-\phihat^{ }_{m\downarrow})$, and $\calU(\phihat^{ }_{m\downarrow})\calU(-\phihat^{ }_{m\downarrow})=1-2\nhat^{ }_{m\uparrow}$.
%%%%%%%%
\section{\label{sec:soft-core} Soft-Core Fermion}
\begin{table}
\caption{\label{tab:soft-hard} Algebra of the soft-core fermions, and its comparison with that of the hard-core bosons. Here, $\left[~,~\right]$ denotes a commutator, and $\left\{~,~\right\}$ denotes an anticommutator.}
\begin{ruledtabular}
\begin{tabular}{ll}
Soft-Core Fermions & Hard-Core Bosons \\ \hline &\\
$\left[\Sigma^-,\Sigma^+\right]=1$ & $\left\{\sigma^-,\sigma^+\right\}=1$ \\ & \\
$\Sigma^z=\left\{\Sigma^+,\Sigma^-\right\}$ & $\sigma^z=\left[\sigma^+,\sigma^- \right]$\\ & \\
$\left[\Sigma^z,\Sigma^\pm\right]=\pm 2\Sigma^\pm$ & $\left[\sigma^z,\sigma^\pm\right]=\pm 2\sigma^\pm$ \\ & \\
Hilbert Space: $\infty$ dimensional & $2$ dimensional \\ & \\
Basis: $\{|n\rangle :~n=0,1,\dots\}$ & $\{|+\rangle, |-\rangle\}$ \\ &\\
$\Sigma^z |n\rangle = (2n+1)|n\rangle$ & $\sigma^z|\pm\rangle=\pm|\pm\rangle$\\ & \\
$\Sigma^+|n\rangle = \sqrt{n+1}|n+1\rangle$ & $\sigma^+|-\rangle=|+\rangle$\\ & \\
$\Sigma^-|0\rangle = 0$ & $\sigma^-|-\rangle=0$\\ & \\
\end{tabular}
\end{ruledtabular}
\end{table}

In this appendix, we introduce a new object, called {\em soft-core fermion}, which is an anticommuting counter-analog of the hard-core boson (the Pauli operators). We define the raising operator $\Sigma^+$ of a soft-core fermion as:
\begin{equation}
\Sigma^+ = \bhat^\dag\phihat
\label{eq:soft-core-fermion}
\end{equation}
where $\bhat^\dag$ is the creation operator of a canonical boson, and $\phihat$ as usual denotes a Majorana fermion such that $\phihat=\chat^\dag+\chat$, where $\chat$ and $\chat^\dag$ are the operators of a canonical fermion. This definition of a soft-core fermion is motivated by the Majorana representation [Eq.~(\ref{eq:inv_rep_sigmaplus})] of $\sigma^+$, the hard-core boson. As the name suggests, a soft-core fermion anticommutes with other fermions and soft-core fermions, while behaving as a boson for itself. This latter property implies no local constraint on the number of such particles (opposite of the hard-core constraint), hence the name soft-core. Also, it commutes with other bosons and hard-core bosons. The algebra of the soft-core fermions is presented in Table~\ref{tab:soft-hard}, and compared with the hard-core bosons.

Interestingly, we can derive a JW type representation for the soft-core fermions, in exactly the same way as we did for the Pauli operators in the previous appendix. We define a unitary transformation: 
\begin{eqnarray}
\calU(\phihat) &=& \exp{ \left[ -i\frac{\pi}{2}(1-\phihat)\nhat \right] } \label{eq:U_phi_2}\\
&=& \left[ \frac{1+(-)^{\nhat} }{2} \right] - \phihat\left[ \frac{1 - (-)^{\nhat}}{2} \right] \nonumber
\end{eqnarray}
such that $\calU^\dag(\phihat)\Sigma^+\calU(\phihat) = \bhat^\dag$. Here, $\nhat=\bhat^\dag\bhat$, and $(-)^{\nhat}=e^{i\pi\nhat}=\cos{(\pi\nhat)}$. If $\nhat$ were the number operator of a fermion, then Eq.~(\ref{eq:U_phi_2}) correctly reduces to Eq.~(\ref{eq:U_phi}). Next, we define an ordered unitary operator $\calU$ for $L$ such soft-core fermions, $\calU=\calU(\phihat_1)\calU(\phihat_2)\cdots\calU(\phihat_L)$. Under $\calU$, the soft-core fermions transform as: 
\begin{eqnarray}
\calU^\dag\Sigma^+_1\calU &=& \bhat^\dag_1 \\
\calU^\dag\Sigma^+_l\calU &=& \left[(-)^{\sum_{m=1}^{l-1}\nhat_m}\right]\bhat^\dag_l \nonumber \\
&& \forall~l=2\dots L
\end{eqnarray}
which is same as the JW transformation, but in terms of the bosons. We have thus presented the mathematical structure of soft-core fermions, whose actual physical usefulness is not obvious at present, but might become so in future.

\bibliography{paper}

\begin{thebibliography}{40}
\bibitem[{fno({\natexlab{b}})}]{fnote_Mattis_Nam}
\bibinfo{note}{It may also be noted that Mattis and Nam~\cite{Mattis_Nam}
  introduced a transformation in which a pair of spin-1/2 objects were combined
  to form another pair of spin-1/2 operators, through a non-linear relation.
  This however does not involve any Fermi variables, and is therefore different
  from our representation.}
\bibitem[{fno({\natexlab{c}})}]{fnote_ph_symmetry}
\bibinfo{note}{One can always add a chemical potential, $U/2$, to make the
  $U$-term particle-hole symmetric. Since the nearest-neighbor hopping is also
  particle-hole symmetric on a bipartite lattice, the Hubbard model is
  invariant under particle-hole transformation, and the chemical potential is
  exactly zero at half-filling. In general, $H$ is not invariant under
  particle-hole transformation, as the hopping does not respect particle-hole
  symmetry on a non-bipartite lattice.}
\bibitem[{fno({\natexlab{d}})}]{fnote_Hubbard1}
\bibinfo{note}{Equation~(\ref{eq:hubbard_new}) can be transformed into a form
  similar to Eq.~(\ref{eq:hubbard_bipartite}) also by performing $\pi/2$-gauge
  transformation on $f_s$'s of, say, $B$-sublattice, and then using the
  identical forms of the new representation on both the sublattices. The
  hopping on a nearest-neighbor bond is now written as:
  $-i\frac{t}{2}\left[\psihat_a\psihat_b + \phihat_a\phihat_b
  \left(\vec{\sigma}\cdot\vec{\tau}\right)\right]$.}
\bibitem[{fno({\natexlab{e}})}]{fnote_anisotropy}
\bibinfo{note}{If the hopping for $\uparrow$ and $\downarrow$ electrons were
  taken to be different ($\tup$ and $\tdown$ respectively), then
  $\vec{\sigma}\cdot\vec{\tau}$ in Eq.~(\ref{eq:hubbard_bipartite}) becomes an
  anisotropic exchange interaction of the form: $\sigma^z\tau^z +
  \frac{\tup}{\tdown}\left(\sigma^x\tau^x + \sigma^y\tau^y\right)$.}
\bibitem[{fno({\natexlab{f}})}]{fnote_Nfermi}
\bibinfo{note}{A representation for the $N$-component fermions, can be
  generated by using a two-component representation, pairwise for $N$ fermion
  operators. For example, the electron operators in a two-orbital problem can
  be written using the two-component representation for spin and orbital
  degrees of freedom successively (in any order).}
\end{thebibliography}
\end{document}